\PassOptionsToPackage{numbers,sort&compress}{natbib}
\documentclass[a4paper,fleqn]{cas-dc}

\usepackage{amsmath,amsfonts,amssymb,bm,mathtools}
\usepackage{booktabs}
\usepackage[ruled,vlined]{algorithm2e}

\usepackage{graphicx}
\usepackage{grffile}
\usepackage{caption}
\usepackage{subcaption}
\usepackage{tikz}
\usetikzlibrary{external}

\usepackage{pgfplots}
\pgfplotsset{compat=newest}
\usepgfplotslibrary{groupplots}
\usepackage{tikzscale}
\usepackage[percent]{overpic}
\usepackage{epstopdf}
\usepackage{import}
\usepackage{placeins}
\usepackage{makecell}
\usepackage{float}
\usepackage{caption}
\usepackage{pgfplots}
\usepackage{balance}
\usepgfplotslibrary{groupplots} 
\pgfplotsset{compat=1.18}       

\newcommand{\vek}[1]{\mathbf{#1}}   
\newcommand{\m}[1]{\mathbf{#1}}      
\newtheorem{definition}{Definition}
\newtheorem{theorem}{Theorem}
\newtheorem{lemma}{Lemma}
\newtheorem{proposition}{Proposition}

\newtheorem{assumption}{Assumption}

\newenvironment{pf}{\par\noindent\textit{Proof: }}{\par\medskip}

\begin{document}
\let\WriteBookmarks\relax
\def\floatpagepagefraction{1}
\def\textpagefraction{.001}

\shorttitle{An $\alpha$-potential linear approximation for LQ differential games}
\shortauthors{B. Varga, et al.}
\title[mode = title]{An $\alpha$-potential linear approximation for feedback Nash equilibria in linear--quadratic differential games}

\fntext[fn1]{The authors contributed equally to this work.}

\author[1]{Balint Varga}[orcid=0000-0002-4189-0105] 
\ead{balint.varga2@kit.edu}
\credit{Conceptualization, Methodology, Formal analysis, Validation, Writing -- original draft, Writing -- review \& editing, Supervision}
\fnmark[1] 
\affiliation[1]{organization={Karlsruhe Institute of Technology (KIT), Institute of Control Systems},
	addressline={Kaiserstra{\ss}e 12},
	city={Karlsruhe},
	postcode={76131},
	country={Germany}}

\author[1]{Gergely L\'or\'ant Kiss}[orcid=0009-0005-3215-2433] 
\credit{Conceptualization, Methodology, Software, Formal analysis, Validation, Visualization, Writing -- review \& editing}
\fnmark[1] 

\cortext[1]{Corresponding author}

\author[2]{Andreas Schwung}[orcid=0000-0001-8405-0977] 
\credit{Methodology, Validation, Writing -- review \& editing}
\affiliation[2]{organization={South Westphalia University of Applied Science, Department of Automation Technology and Learning Systems},
	addressline={Lübecker Ring 2},
	city={Soest},
	postcode={59494},
	country={Germany}}

\begin{abstract}
	Feedback Nash equilibria of infinite-horizon linear--quadratic (LQ) differential games are characterized by coupled algebraic Riccati equations (cAREs), whose numerical solution generally relies on iterative algorithms and can be highly sensitive to their initialization. Motivated by the $\alpha$-potential-game framework, this letter proposes an $\alpha$\emph{-potential linear ($\alpha$PL) approximation} for constructing a systematic, game-dependent initialization. For an $N$-player LQ game, a reference potential game is constructed. Its equilibrium requires solving only one standard algebraic Riccati equation. The original cAREs are subsequently linearized around this reference solution, yielding a coupled linear system whose solution provides player-specific corrections and hence an approximate feedback Nash equilibrium. The resulting approximation is used to initialize three standard cARE solvers. Numerical experiments on $10^4$ randomly generated two-player games show that the proposed initialization reduces the mean number of iterations. Further experiments for $N=2,\ldots,13$ show that the reduction in iteration count persists as the number of players increases.
\end{abstract}


\begin{keywords}
	Linear--quadratic differential games, coupled algebraic Riccati equations, feedback Nash equilibrium, potential games, numerical initialization.
\end{keywords}

\maketitle

\section{Introduction}\label{sec:intro}

Linear-quadratic (LQ) differential games provide a tracable framework for dynamic decision problems involving multiple interacting agents \cite{basar2018,engwerda2005,thoemmes_2026_IFAC}. A central solution concept is the feedback Nash equilibrium, which, for infinite-horizon LQ games, is characterized by a set of coupled algebraic Riccati equations (cAREs) \cite{engwerda2005,weeren1999}. Unlike the standard single-player ARE, these coupled equations admit no closed-form solution and must be solved iteratively \cite{engwerda2007}. This is relevant in real-time human--machine systems, where changing interaction conditions or human preferences may require recomputing the game and its equilibrium online \cite{franceschi2024,HVROC_2023_Kille,varga2024}. In such settings, efficient, reliable cARE solvers are essential, and their convergence and cost can depend strongly on initialization.

Potential games offer a useful structural simplification for differential games, as their feedback Nash equilibrium can be characterized through the optimum of a common potential function \cite{monderer1996,fonseca2018}. Successful applications of potential games range from distributed
self-optimization in modular production systems \cite{schwung2022} to
game-theoretic analyses of social networks, where coevolutionary dynamics of
actions and opinions admit a potential representation \cite{liang2026}. In the LQ setting, an exact potential structure converts the equilibrium problem into a standard optimal-control problem that can be solved through a single ARE. Such a structure, however, is only available under restrictive conditions.  
The $\alpha$-potential-game framework relaxes this requirement by allowing the change in a player's objective under a unilateral deviation to differ from the corresponding change in a potential function by a bounded amount \cite{Guo_Alpha_2025,Guo_2025}. This motivates exploiting an approximate potential structure also for LQ games that are not exact potential games.

Based on this idea, we propose an $\alpha$\emph{-potential linear ($\alpha$PL) approximation} for initializing the solver of cAREs of an $N$-player LQ differential game. A reference potential game is constructed by aggregating the players' input matrices and state penalties, and its solution is obtained from a single standard ARE. The original cAREs are then linearized around this reference solution. Neglecting second-order terms yields a coupled linear system whose solution provides player-specific corrections and an approximate feedback Nash equilibrium. The main contributions of this letter are:
\begin{itemize}
	\item An $\alpha$-potential reference game systematically initializes iterative solvers for coupled Riccati equations.
	\item The initializer solves one ARE and one linear system.
	\item Invalid-solution rates, number of iterations, and total computation time are decreased with the proposed initializer.
\end{itemize}

\section{Problem Formulation}
\label{sec:problem}

We consider the $N$-player infinite-horizon LQ differential game with players $\Gamma:=\{1,\dots,N\}$, defined as follows.
\begin{definition}
\upshape (\textbf{$N$-player infinite-horizon LQ differential game})
\label{def:game}
An $N$-player infinite-horizon linear--quadratic (LQ) differential game is a tuple 
$$
G=\big(\Gamma,\,\m{A},\,(\m{B}_i)_{i\in\Gamma},\,(\m{Q}_i)_{i\in\Gamma},\,(\m{R}_{ij})_{i,j\in\Gamma}\big)
$$ 
with state $\vek{x}(t)\in\mathbb{R}^n$ and control inputs $\vek{u}_i(t)\in\mathbb{R}^{m_i}$, whose dynamics are
\begin{equation}
	\dot{\vek{x}}(t) = \m{A}\,\vek{x}(t) + \sum_{j=1}^{N}\m{B}_j\vek{u}_j(t),
	\qquad \vek{x}(0)=\vek{x}_0,
	\label{eq:dyn}
\end{equation}
and in which each Player $i\in\Gamma$ minimizes the cost functional
\begin{equation}
	J_i = \tfrac{1}{2}\!\int_{0}^{\infty}\!\Big(\vek{x}(t)^\top\m{Q}_i\vek{x}(t)	
	+ \sum_{j=1}^{N}\vek{u}_j(t)^\top\m{R}_{ij}\vek{u}_j(t)\Big)\,\mathrm{d}t .
	\label{eq:cost}
\end{equation}
\end{definition}

\begin{assumption}[Standing regularity conditions]
\label{ass:reg}
For all $i,j\in\Gamma$, the cost matrices satisfy $\m{Q}_i=\m{Q}_i^\top\succeq \m{0}$, $\m{R}_{ij}=\m{R}_{ij}^\top$, and $\m{R}_{ii}\succ \m{0}$. Moreover, each pair $(\m{A},\m{B}_i)$ is stabilizable and each pair $(\m{Q}_i^{1/2},\m{A})$ is detectable.
\end{assumption}
In the LQ setting the feedback Nash equilibrium is attained by stationary linear
policies $\vek{u}_i^*=-\m{K}_i\vek{x},$ see~\cite{weeren1999,engwerda2005}. Writing the
value function of each player as $V_i^*(\vek{x})=\tfrac12\vek{x}^\top\m{P}_i\vek{x}$
and introducing
$$
\m{S}_i := \m{B}_i\m{R}_{ii}^{-1}\m{B}_i^\top \; \mathrm{and}	\; \m{S}_{ij} := \m{B}_j\m{R}_{jj}^{-1}\m{R}_{ij}\m{R}_{jj}^{-1}\m{B}_j^\top,
$$ the equilibrium is characterized as follows.

\begin{proposition} 
\upshape \textbf{(Riccati characterization of the feedback Nash equilibrium~\cite{engwerda2005}).}
\label{prop:care}
Let Definition~\ref{def:game} and Assumption~\ref{ass:reg} hold. A profile of
stationary linear policies $\vek{u}_i^*=-\m{K}_i\vek{x}$ constitutes a feedback
Nash equilibrium with value matrices $\m{P}_i=\m{P}_i^\top\succ\m{0}$ if the
$\m{P}_i$ solve the coupled algebraic Riccati equations (cAREs)
\begin{equation}
	\begin{split}
		0 = &\Big(\m{A}-\!\sum_{j=1}^{N}\m{S}_j\m{P}_j\Big)^{\!\top}\!\m{P}_i
		+ \m{P}_i\Big(\m{A}-\!\sum_{j=1}^{N}\m{S}_j\m{P}_j\Big)\\
		&+ \sum_{j=1}^{N}\m{P}_j\m{S}_{ij}\m{P}_j + \m{Q}_i,
		\qquad \forall i\in\Gamma,
	\end{split}
	\label{eq:cARE}
\end{equation}
and the associated closed-loop matrix $\m{A}_{cl}:=\m{A}-\sum_{i=1}^{N}\m{S}_i\m{P}_i$ is Hurwitz. The equilibrium feedback gains are then $\m{K}_i=\m{R}_{ii}^{-1}\m{B}_i^\top\m{P}_i$.
\end{proposition}

Since \eqref{eq:cARE} has no closed-form solution, the matrices $\m{P}_i$ are obtained iteratively. This letter considers three iterative schemes, referred to for brevity as \emph{policy iteration}, \emph{sequential iteration}, and \emph{Newton--Raphson iteration}. The first corresponds to \cite{li2019} and updates the player matrices by solving decoupled Lyapunov equations at each iteration. The sequential iteration corresponds to Algorithm~6 of \cite{engwerda2007} and successively solves stabilizing single-player AREs using the most recently updated player matrices. Finally, the Newton--Raphson iteration corresponds to Algorithm~7 of \cite{engwerda2007} and jointly updates the coupled system through linear matrix equations. All three require an initial guess $\m{P}_i^0$ such that $\m{A}-\sum_i\m{S}_i\m{P}_i^0$ is Hurwitz, and their convergence hinges on this choice. Two initialization strategies are considered as baselines: the game-agnostic identity $\m{P}_i^0=\m{I}$ and the constructive initialization of \cite{engwerda2007}, which sequentially solves decoupled single-player AREs. We denote these by $\m{P}^0_{\m{I}}$ and $\m{P}^0_{\mathrm{engw}}$, respectively, and the proposed initialization by $\m{P}^0_{\alpha\mathrm{PL}}$.

\section{An $\alpha$-Potential linear approximation}\label{sec:method}
\subsection{The $\alpha$-Potential Game Structure}
The proposed method is motivated by the framework of $\alpha$\emph{-potential games} \cite{Guo_Alpha_2025,Guo_2025}, formalized as follows for the differential game of Definition~\ref{def:game}. Let $\mathcal{U}_i$ denote the admissible feedback strategies of Player~$i$ and $\vek{u}(t)=(\vek{u}_i(t),\vek{u}_{-i}(t))$ a strategy profile.

\begin{definition}[$\alpha$-potential game]
\label{def:alpha_potential}
The game $G$ of Definition~\ref{def:game} is an \emph{$\alpha$-potential game} if there exist a scalar $\alpha\geq0$ and a potential functional $J^{(p)}:\mathcal{U}_1\times\cdots\times\mathcal{U}_N\to\mathbb{R}$ such that, for every player $i\in\Gamma$, every pair $\vek{u}_i(t),\vek{u}_i'(t)\in\mathcal{U}_i$, and every fixed $\vek{u}_{-i}(t)$,
\begin{equation}
	\begin{split}
		\Big|&\left(J_i(\vek{u}_i'(t),\vek{u}_{-i}(t)) - J_i(\vek{u}_i(t),\vek{u}_{-i}(t)) \right)\\
		&-\left( J^{(p)}(\vek{u}_i'(t),\vek{u}_{-i}(t)) - J^{(p)}(\vek{u}_i(t),\vek{u}_{-i}(t)) \right) \Big|
		\leq \alpha .
	\end{split}
	\label{eq:alpha_potential}
\end{equation}
For $\alpha=0$ the game is an \emph{exact potential game}; for $\alpha>0$ the parameter $\alpha$ bounds the admissible difference between the unilateral variation of $J_i$ and that of the potential $J^{(p)}$.
\end{definition}

Because the reference functional $J^{(p)}$ shares the control penalties of each $J_i$, the deviation gap in \eqref{eq:alpha_potential} reduces to a state-weighted quadratic in the residual $\Delta\m{Q}_i:=\m{Q}_i-\m{Q}$, so the game \eqref{eq:dyn}, \eqref{eq:cost} is $\alpha$-potential with respect to $J^{(p)}$ with $\alpha\le\gamma\max_{i\in\Gamma}\|\Delta\m{Q}_i\|_2$ over admissible strategies of bounded trajectory energy $\gamma$, and $\alpha=0$ exactly when $\m{Q}_i=\m{Q} \ \forall \, i$. The same residual $\Delta\m{Q}_i$ drives the linearization of Section~\ref{sec:lin} and the error bound of Lemma~\ref{lem:error}, linking the objective-level parameter $\alpha$ to the cARE-level deviation.

In this work, our objective is not to evaluate the smallest such $\alpha$, but to exploit this proximity-to-potential principle computationally. We therefore construct $J^{(p)}$ as an exact-potential reference and linearize the original cAREs about its Riccati solution; see Subsections~\ref{sec:ref_pot_game_gen} and Section~\ref{sec:lin}. In this construction, $\alpha$ measures deviation at the objective-functional level, whereas the player-specific Riccati deviations and the resulting second-order remainder measure deviation at the cARE level. 

\subsection{Construction of the Reference $\alpha$-Potential Game} \label{sec:ref_pot_game_gen}
\begin{definition}[Reference Potential System]
\label{def:decoupled_reference_potential}
Assume that \\ the control penalties are decoupled across players, i.e., $\m{R}_{ij} = \m{0}$ and consequently $\m{S}_{ij} = \m{0}$ for all $i \neq j$, such that the coupling sum $\sum_{j=1}^{N}\m{P}_j\m{S}_{ij}\m{P}_j$ in the cARE \eqref{eq:cARE} reduces to $\m{P}_i\m{S}_i\m{P}_i$.
Let $\vek{u} := \mathrm{col}(\vek{u}_1,\ldots,\vek{u}_N)$ denote the aggregate input vector and $\m{B} := \big[\m{B}_1\ \cdots\ \m{B}_N\big]$ the corresponding input matrix. The \emph{reference potential functional} $J^{(p)}(\vek{u})$ associated with system \eqref{eq:dyn} is defined as the LQ cost functional
\begin{equation}
	J^{(p)}
	:=
	\frac{1}{2}\int_{0}^{\infty}
	\left(
	\vek{x}^{\top}(t)\m{Q}\vek{x}(t)
	+
	\vek{u}^{\top}(t)\m{R}\vek{u}(t)
	\right)\mathrm{d}t,
	\label{eq:potential_cost}
\end{equation}
which defines the exact potential reference problem used for the approximation in the $\alpha$-potential relation \eqref{eq:alpha_potential}\footnote{For the sake of notational simplicity, we omit the index $[ ]^p$ of the matrices of the differential potential game.}.
\end{definition}

\begin{definition}[Reference cost aggregation]
\label{def:aggregate}
Given weights
\begingroup
\setlength{\abovedisplayskip}{2pt}
\setlength{\abovedisplayshortskip}{2pt}
\begin{equation} \label{eq:weights_calculation}
	\theta_i= \frac{1}{N} \forall i\in\Gamma,
\end{equation}
\endgroup
the reference penalty matrices are
\begin{equation}
		\m{Q}:=\sum_{i=1}^{N}\theta_i\,\m{Q}_i, \;
		\m{R}:=\mathrm{diag}(\m{R}_{11},\dots,\m{R}_{NN}).
  \label{eq:QRaggregate}
\end{equation}
\end{definition}

\begin{proposition}[Well-posed potential reference game]
\label{prop:reference}
Let Assumption~\ref{ass:reg} hold, suppose that $\m{R}_{ij}=\m{0}$ for $i\neq j$, and let $\m{Q}$ and $\m{R}$ be defined by \eqref{eq:QRaggregate}. Then:
\begin{enumerate}
    \item[(i)] $\m{Q}=\m{Q}^{\top}\succeq\m{0}$ and
    $\m{R}=\m{R}^{\top}\succ\m{0}$;
    \item[(ii)] assigning the common objective $J^{(p)}$ to all players defines an exact potential reference game with potential $J^{(p)}$;
    \item[(iii)] with
    \begin{equation}
        \m{S}:=\m{B}\m{R}^{-1}\m{B}^{\top}
        =\sum_{j=1}^{N}\m{B}_j\m{R}_{jj}^{-1}\m{B}_j^{\top}
        =\sum_{j=1}^{N}\m{S}_j,
        \label{eq:Saggregate}
    \end{equation}
    the reference ARE
    \begin{equation}
        \m{0}=\m{A}^{\top}\m{P}+\m{P}\m{A}
        -\m{P}\m{S}\m{P}+\m{Q}
        \label{eq:ARE_p}
    \end{equation}
    has a unique stabilizing symmetric solution $\m{P}\succeq\m{0}$, and $\m{A}_{cl}:=\m{A}-\m{S}\m{P}$ is Hurwitz.
\end{enumerate}
\end{proposition}

\begin{pf}
Part (i) follows because $\m{Q}$ is the average of symmetric positive-semidefinite matrices and $\m{R}$ is block diagonal with positive-definite diagonal blocks.

For (ii), let every player in the reference game have cost $J^{(p)}$. Then every unilateral variation of a player's cost equals the corresponding variation of $J^{(p)}$, so Definition~\ref{def:alpha_potential} holds with $\alpha=0$.

Finally, $(\m{A},\m{B})$ is stabilizable because each $(\m{A},\m{B}_i)$ is stabilizable. Moreover, $(\m{Q}^{1/2},\m{A})$ is detectable: if an unstable mode is unobservable through $\m{Q}^{1/2}$, then
\[
    0=\vek{v}^{*}\m{Q}\vek{v}
    =\frac{1}{N}\sum_{i=1}^{N}\vek{v}^{*}\m{Q}_i\vek{v}
\]
implies $\m{Q}_i^{1/2}\vek{v}=\m{0}$ for every $i$, contradicting Assumption~\ref{ass:reg}. Since $\m{R}\succ\m{0}$, standard LQ theory therefore guarantees the unique stabilizing solution of \eqref{eq:ARE_p} and the Hurwitz property of $\m{A}_{cl}$, see also \cite{laub1979,vandooren1981}.
\qed
\end{pf}

\section{Linearization of the cAREs using the $\alpha$-potential reference}
\label{sec:lin}

After finding the reference solution $\m{P}$, we linearize the cAREs \eqref{eq:cARE} around it. The resulting linear system is solved for the player-specific deviations $\Delta\m{P}_i$ and yields an approximate feedback Nash equilibrium, which is used to initialize the iterative solvers. 

\subsection{Proposed First Order Linearization}
We now state the proposed linearization. To obtain a coupled system of linear equations, we expand the cAREs around the reference solution $\m{P}$ and neglect all terms that are second order in the player-specific deviations. Although we initially drop these terms for tractability, we do not discard them without justification: Section~\ref{sec:error} provides a formal error analysis showing that the resulting approximation error is second order in the deviation magnitude.

\begin{theorem}[$\alpha$-potential linear approximation]
	\label{thm:apl}
	Consider the $N$-player LQ game \eqref{eq:dyn}--\eqref{eq:cost} with decoupled
	control penalties, $\m{R}_{ij}=\m{0}$ for $i\neq j$, let $\m{P}$ be the
	stabilizing solution of the reference ARE \eqref{eq:ARE_p}, and set
	$\m{A}_{cl}=\m{A}-\m{S}\m{P}$. Expand each equilibrium value matrix and state
	penalty about the reference solution,
	\begin{equation}
		\m{P}_i = \m{P} + \Delta\m{P}_i,\qquad
		\m{Q}_i = \m{Q} + \Delta\m{Q}_i,\qquad \forall i\in\Gamma,
		\label{eq:dev}
	\end{equation}
	so that $\m{P}$ comprises the potential structure and $\Delta\m{P}_i$ the
	player-specific deviation. Neglecting in the cARE \eqref{eq:cARE} the terms that
	are second order in the deviations $\{\Delta\m{P}_i\}_{i\in\Gamma}$ linearizes it
	into a system of $N\!\cdot\!n^2$ linear equations in the $\Delta\m{P}_i$.
	Whenever the solution of this system of linear equations is nonsingular, the system admits
	a unique solution $\{\Delta\m{P}_i^{\mathrm{approx}}\}_{i\in\Gamma}$, and the $\alpha$PL approximation
	\begin{equation}
		\m{P}_i^{\mathrm{approx}} = \m{P} + \Delta\m{P}_i^{\mathrm{approx}}, \qquad \forall i\in\Gamma,
		\label{eq:Papprox}
	\end{equation}
	is well-defined and computable by a single linear solver.
\end{theorem}

\begin{pf}
	Under the decoupling assumption the cARE \eqref{eq:cARE} for Player~$i$ reduces to
	\begin{equation}
		0 = \Big(\m{A}-\!\!\sum_{\substack{j=1\\ j\neq i}}^{N}\m{S}_j\m{P}_j\Big)^{\!\top}\!\m{P}_i
		+ \m{P}_i\Big(\m{A}-\!\!\sum_{\substack{j=1\\ j\neq i}}^{N}\m{S}_j\m{P}_j\Big)
		- \m{P}_i\m{S}_i\m{P}_i + \m{Q}_i.
		\label{eq:ARE_i}
	\end{equation}
	Substituting \eqref{eq:dev} into \eqref{eq:ARE_i} and using the symmetry of
	$\m{P}$ and $\m{S}_j$, the terms group by order in the deviations. Collecting the
	second-order contributions into
	\begin{equation}
		\varepsilon_i := \!\!\sum_{\substack{j=1\\ j\neq i}}^{N}\!\!\big(\Delta\m{P}_j\m{S}_j\Delta\m{P}_i
		+ \Delta\m{P}_i\m{S}_j\Delta\m{P}_j\big) + \Delta\m{P}_i\m{S}_i\Delta\m{P}_i,
		\label{eq:eps}
	\end{equation}
	and consolidating the first-order terms in $\Delta\m{P}_i$ via
	$\m{A}^\top\Delta\m{P}_i-\sum_{j}\m{P}\m{S}_j\Delta\m{P}_i=\m{A}_{cl}^\top\Delta\m{P}_i$,
	equation \eqref{eq:ARE_i} separates as
	\begin{equation}
		\begin{split}
			0 &= \underbrace{\m{A}^\top\m{P} + \m{P}\m{A} - \m{P}\m{S}\m{P} + \m{Q}}_{=\,\m{0}\ \text{by \eqref{eq:ARE_p}}}\\
			&- \sum_{\substack{j=1\\ j\neq i}}^{N}\m{P}\m{S}_j\m{P} + \Delta\m{Q}_i
			+ \m{A}_{cl}^\top\Delta\m{P}_i + \Delta\m{P}_i\m{A}_{cl}\\
			&- \sum_{\substack{j=1\\ j\neq i}}^{N}\!\big(\Delta\m{P}_j\m{S}_j\m{P} + \m{P}\m{S}_j\Delta\m{P}_j\big)
			- \varepsilon_i .
		\end{split}
		\label{eq:grouped}
	\end{equation}
	The underbraced group vanishes by the reference ARE \eqref{eq:ARE_p}. Retaining
	terms up to first order in the player-specific deviations, i.e.\ neglecting the
	second-order remainder $\varepsilon_i$, leaves the coupled linear matrix equation
	\begin{equation}
		\begin{split}
			\m{A}_{cl}^\top\Delta\m{P}_i &+ \Delta\m{P}_i\m{A}_{cl}
			- \!\!\sum_{\substack{j=1\\ j\neq i}}^{N}\!\!
			\big(\Delta\m{P}_j\m{S}_j\m{P}
			+ \m{P}\m{S}_j\Delta\m{P}_j\big)\\
			&= -\Delta\m{Q}_i
			+ \!\!\sum_{\substack{j=1\\ j\neq i}}^{N}
			\m{P}\m{S}_j\m{P}.
		\end{split}
		\label{eq:lin}
	\end{equation}
	Applying the vectorization identity
	$\mathrm{vec}(\m{A}\m{X}\m{B})=(\m{B}^\top\otimes\m{A})\,\mathrm{vec}(\m{X})$,
	with $\otimes$ the Kronecker product, \eqref{eq:lin} becomes, for each $i$,
	\begin{equation}
		\m{L}_{cl}\,\vek{p}_i - \!\!\sum_{\substack{j=1\\ j\neq i}}^{N}\m{K}_{ij}\,\vek{p}_j = \vek{b}_i,
		\label{eq:vec_i}
	\end{equation}
	where $\m{L}_{cl} = \m{I}\otimes\m{A}_{cl}^\top + \m{A}_{cl}^\top\otimes\m{I}$,
	$\; \m{K}_{ij} = (\m{S}_j\m{P})^\top\otimes\m{I} + \m{I}\otimes\m{P}\m{S}_j$,
	$\;\vek{p}_i = \mathrm{vec}(\Delta\m{P}_i^{\mathrm{approx}})$, and
	$\;\vek{b}_i = \mathrm{vec}\Big(-\Delta\m{Q}_i + \sum_{\substack{j=1\\ j\neq i}}^{N}\m{P}\m{S}_j\m{P}\Big)$,
	with $\vek{p}_i,\vek{b}_i\in\mathbb{R}^{n^2}$ and
	$\m{L}_{cl},\m{K}_{ij}\in\mathbb{R}^{n^2\times n^2}$. Stacking \eqref{eq:vec_i}
	over all players $i\in\Gamma$ yields the block linear system
	\begin{equation}
		\begin{bmatrix}
			\m{L}_{cl} & -\m{K}_{12} & \cdots & -\m{K}_{1N}\\
			-\m{K}_{21} & \m{L}_{cl} & \cdots & -\m{K}_{2N}\\
			\vdots & \vdots & \ddots & \vdots\\
			-\m{K}_{N1} & -\m{K}_{N2} & \cdots & \m{L}_{cl}
		\end{bmatrix}
		\begin{bmatrix}\vek{p}_1\\ \vek{p}_2\\ \vdots\\ \vek{p}_N\end{bmatrix}
		=
		\begin{bmatrix}\vek{b}_1\\ \vek{b}_2\\ \vdots\\ \vek{b}_N\end{bmatrix}
		\label{eq:linsys}
	\end{equation}
	of $N\!\cdot\!n^2$ equations. When its coefficient matrix is nonsingular, the
	solution $\vek{p}_i=\mathrm{vec}(\Delta\m{P}_i^{\mathrm{approx}})$ is unique;
	reshaping it into $\Delta\m{P}_i^{\mathrm{approx}}$ and adding $\m{P}$ gives the
	approximation \eqref{eq:Papprox}. \qed
\end{pf}

\subsection{Approximation Error Analysis}
\label{sec:error}

Let $\mathcal{F}_i(\m{P}_1,\ldots,\m{P}_N)$ denote the left-hand side of the
$c$ARE \eqref{eq:ARE_i}. Since \eqref{eq:lin} cancels all constant and
first-order terms in \eqref{eq:grouped}, the residual of the $\alpha$PL
approximation \eqref{eq:Papprox} is given exactly by the neglected
second-order terms, $$
\mathcal{F}_i
\big(\m{P}_1^{\mathrm{approx}},\ldots,\m{P}_N^{\mathrm{approx}}\big)=
-\varepsilon_i^{\mathrm{approx}},$$
where $\varepsilon_i^{\mathrm{approx}}$ is obtained from \eqref{eq:eps} by
replacing $\Delta\m{P}_j$ with
$\Delta\m{P}_j^{\mathrm{approx}}$. Thus, the cARE residual is second order in
the correction magnitude.

\begin{algorithm}[!t]
	\caption{$\alpha$-Potential Linear ($\alpha$PL) Approximation}
	\label{alg:npl}
	\DontPrintSemicolon
	\KwIn{$N$-player LQ game \eqref{eq:dyn}--\eqref{eq:cost}.}
	Compute the weights by \eqref{eq:weights_calculation}.\;
	Build the reference game via \eqref{eq:QRaggregate} and \eqref{eq:Saggregate}.\;
	Solve the ARE \eqref{eq:ARE_p} for $\m{P}$.\;
	Assemble and solve the linear system \eqref{eq:linsys} for $\vek{p}_i$.\;
	Reshape $\vek{p}_i\!\to\!\Delta\m{P}_i^{\mathrm{approx}}$ and set $\m{P}_i^{\mathrm{approx}}=\m{P}+\Delta\m{P}_i^{\mathrm{approx}}$.\;
	\KwOut{Use $\m{P}_i^{\mathrm{approx}}$ as the initial point of an iterative cARE solver.}
\end{algorithm}

\begin{lemma}[Error of the $\alpha$PL approximation]
	\label{lem:error}
	Let $\m{M}$ denote the block coefficient matrix in \eqref{eq:linsys}, assume that $\m{M}$ is nonsingular, and let 
	$$
	\vek{p}^{\alpha} := \mathrm{col}\big(\mathrm{vec}(\Delta\m{P}_1^{\mathrm{approx}}),\ldots, \mathrm{vec}(\Delta\m{P}_N^{\mathrm{approx}})\big)
	$$ 
	denote the $\alpha$PL solution. Furthermore, let 
	$$
	\vek{p}^{*} := \mathrm{col}\big(\mathrm{vec}(\Delta\m{P}_1^{*}),\ldots, \mathrm{vec}(\Delta\m{P}_N^{*})\big)
	$$
	correspond to an exact cARE solution $\m{P}_i^{*}=\m{P}+\Delta\m{P}_i^{*}$. Defining $\delta:=\max_{i\in\Gamma}\|\Delta\m{P}_i^{*}\|_F$ and $s:=\max_{i\in\Gamma}\|\m{S}_i\|_2$, the approximation error satisfies
	\begin{equation}
		\big\|\vek{p}^{*}-\vek{p}^{\alpha}\big\|_2 \le \sqrt{N}\,(2N-1)\,\|\m{M}^{-1}\|_2\,s\,\delta^2 .
		\label{eq:errbound}
	\end{equation}
	For a uniformly bounded inverse $\m{M}^{-1}$, the $\alpha$PL approximation error is second order in the deviation~magnitude~$\delta$.
\end{lemma}

\begin{pf}
	Substituting $\m{P}_i^{*}=\m{P}+\Delta\m{P}_i^{*}$ into \eqref{eq:ARE_i} without neglecting the second-order terms and vectorizing gives $\m{M}\vek{p}^{*} = \vek{b}+\boldsymbol{\varepsilon}^{*}$, where $\vek{b}$ is the right-hand side of \eqref{eq:linsys} and $\boldsymbol{\varepsilon}^{*}$ stacks $\mathrm{vec}(\varepsilon_i^{*})$, with
	\[
	\varepsilon_i^{*} = \sum_{\substack{j=1\\j\neq i}}^{N} \left( \Delta\m{P}_j^{*}\m{S}_j\Delta\m{P}_i^{*} + \Delta\m{P}_i^{*}\m{S}_j\Delta\m{P}_j^{*} \right) + \Delta\m{P}_i^{*}\m{S}_i\Delta\m{P}_i^{*}.
	\]
	In contrast, the $\alpha$PL approximation satisfies $\m{M}\vek{p}^{\alpha}=\vek{b}$. Subtracting both systems yields
	\begin{equation}
		\vek{p}^{*}-\vek{p}^{\alpha} = \m{M}^{-1}\boldsymbol{\varepsilon}^{*}.
		\label{eq:error_identity}
	\end{equation}
	By submultiplicativity and the definitions of $s$ and $\delta$, $\|\varepsilon_i^{*}\|_F \le (2N-1)s\delta^2$, and therefore $\|\boldsymbol{\varepsilon}^{*}\|_2 \le \sqrt{N}\,(2N-1)s\delta^2$. Combining this inequality with \eqref{eq:error_identity} gives \eqref{eq:errbound}.
\end{pf}

\textit{Remark:} Since the stabilizing solution of \eqref{eq:ARE_p} makes $\m{A}_{cl}$ Hurwitz, the Lyapunov block $\m{L}_{cl}$ is nonsingular. A sufficient condition for $\m{M}$ to be nonsingular is
\begin{equation}
	\max_i\sum_{j\neq i}
	\|\m{L}_{cl}^{-1}\m{K}_{ij}\|_2<1 .
	\label{eq:invertibility}
\end{equation}
Hence, the linearized problem is well posed whenever the coupling is sufficiently weak relative to the reference Lyapunov operator. The overall procedure is summarized in Algorithm~\ref{alg:npl}.

\section{Numerical results}\label{sec:results}
The proposed method is evaluated and compared with the two other solvers presented in Section~\ref{sec:problem}: The identity ($\m{P}^0_{\m{I}}$) and the decoupled-ARE ($\m{P}^0_{\mathrm{engw}}$) strategies. Random games are generated as follows. The system matrix $\m{A}$ and the input matrices $\m{B}_i$ are sampled entrywise, independently and uniformly, from $(-20,20)$; the pair $(\m{A},[\m{B}_1\cdots\m{B}_N])$ is then tested for stabilizability and resampled until this holds. 

The symmetric matrices $\m{Q}_i$ and $\m{R}_i$, however, require a more careful construction, as the prescribed definiteness properties must be satisfied by design. Any real symmetric matrix $\m{M} \in \mathbb{R}^{n \times n}$ admits a decomposition of the form
\begin{equation}
    \m{M} = \m{V} \m{\Lambda} \m{V}^\top,
    \label{eq:matrix_generation}
\end{equation}
where $\m{V} \in \mathbb{R}^{n \times n}$ is an orthonormal matrix and $\m{\Lambda} = \mathrm{diag}(\lambda_i)$ is a diagonal matrix whose entries are the eigenvalues of $\m{M}$, see e.g.~Theorem~2.27 in~\cite{engwerda2005}.
Furthermore, by Theorem~2.28 in~\cite{engwerda2005}, $\m{M}$ is positive definite (semidefinite) if and only if all eigenvalues satisfy $\lambda_i > 0$ ($\lambda_i \geq 0$).

Based on these two theorems (Theorem~2.27 and~2.28 in~\cite{engwerda2005}) , the symmetric matrices $\m{Q}_i$ and $\m{R}_i$ are constructed as follows.
A diagonal matrix $\m{\Lambda} \in \mathbb{R}^{n \times n}$ is formed with eigenvalues drawn independently and uniformly from $[0,\, 20)$ for positive semidefinite matrices (e.g. $\m{Q}_i$), or from $(0,\, \alpha)$ for positive definite matrices (e.g. $\m{R}_i$).
A random orthogonal matrix $\m{V} \in \mathbb{R}^{n \times n}$ is sampled uniformly from the orthogonal group $O(n)$, and the matrix is assembled according to~\eqref{eq:matrix_generation}.
This construction guarantees by design that the resulting matrix is symmetric and has the prescribed definiteness property.
For the positive definite matrices $\m{R}_i$, a small regularization term $\varepsilon \m{I}$ with $\varepsilon = 10^{-3}$ is added to ensure strict positive definiteness in the presence of numerical errors. 

The common bound of $20$ keeps all matrices at comparable magnitude, preventing degenerate cost functionals in which a single state or input term dominates. Unless stated otherwise, the study uses $10^4$ instances with ${N=n=m_i=2}$. 

A solution is said to be \emph{invalid} if the solver (a)~does not converge within $500$ iterations (tolerance $10^{-8}$), (b)~returns a non-symmetric $\m{P}^{\mathrm{approx}}_i$, or (c)~violates the closed-loop stability (admissibility) condition. Accuracy is measured by the Frobenius norm of the cARE residual, which is the left-hand side of \eqref{eq:cARE} evaluated at the computed $\m{P}_i$. 

We report five aspects in turn: a) the rate of valid solutions, b) the solution accuracy, c) the convergence speed and computation time, d) the uniqueness of the identified equilibrium, and e) the scalability in the number of players.

\begin{table*}[t]
\footnotesize
\begin{minipage}[t]{0.32\textwidth}
	\raggedright
	\caption{Invalid-solution rate (\%) \\
		over $10^4$ games ($N=n=m_i=2$).}
	\label{tab:invalid}
	\begin{tabular}{@{}lccc@{}}
		\toprule
		Solver & $\m{P}^0_{\m{I}}$ & $\m{P}^0_{\mathrm{engw}}$ & $\m{P}^0_{\alpha\mathrm{PL}}$ \\
		\midrule
		\makecell[l]{Policy\\ iteration} & $6.57$ & $0.04$ & $\mathbf{0.04}$ \\
		Sequential                       & $0.01$ & $0.02$ & $\mathbf{0.01}$ \\
		\makecell[l]{Newton--\\ Raphson}  & $7.90$ & $3.79$ & $\mathbf{1.48}$ \\
		\bottomrule
	\end{tabular}
\end{minipage}%
\hfill
\begin{minipage}[t]{0.32\textwidth}
	\raggedright
	\caption{Mean number of iterations \\
		until convergence.}
	\label{tab:iter}
	\begin{tabular}{@{}lccc@{}}
		\toprule
		Solver & $\m{P}^0_{\m{I}}$ & $\m{P}^0_{\mathrm{engw}}$ & $\m{P}^0_{\alpha\mathrm{PL}}$ \\
		\midrule
		\makecell[l]{Policy\\ iteration} & $24.30$ & $23.17$ & $\mathbf{20.93}$ \\
		Sequential                       & $12.54$ & $12.05$ & $\mathbf{11.08}$ \\
		\makecell[l]{Newton--\\ Raphson}  & $\phantom{0}7.44$ & $\phantom{0}5.93$ & $\phantom{0}\mathbf{4.38}$ \\
		\bottomrule
	\end{tabular}
\end{minipage}%
\hfill
\begin{minipage}[t]{0.32\textwidth}
	\raggedright
	\caption{Mean total computation time (ms), \\
		initialization plus solver.}
	\label{tab:time}
	\begin{tabular}{@{}lccc@{}}
		\toprule
		Solver & $\m{P}^0_{\m{I}}$ & $\m{P}^0_{\mathrm{engw}}$ & $\m{P}^0_{\alpha\mathrm{PL}}$ \\
		\midrule
		\makecell[l]{Policy\\ iteration} & $\mathbf{2.606}$ & $3.104$ & $2.651$ \\
		Sequential                       & $7.004$ & $7.287$ & $\mathbf{6.498}$ \\
		\makecell[l]{Newton--\\ Raphson}  & $1.103$ & $1.539$ & $\mathbf{1.090}$ \\
		\bottomrule
	\end{tabular}
\end{minipage}
\end{table*}

\newpage
\subsection{Rate of valid solutions}
Table~\ref{tab:invalid} shows that the sequential solver is essentially always valid, whereas policy iteration and, more noticeably, Newton--Raphson can fail. Resolving the invalid cases into their causes reveals that non-symmetry of $\m{P}_i$ never occurs; the failures are due exclusively to non-convergence and to violations of the admissibility condition. Policy iteration fails almost only under the game-agnostic $\m{P}^0_{\m{I}}$, while Newton-Raphson is the most initialization-sensitive: $\m{P}^0_{\alpha\mathrm{PL}}$ lowers its invalid rate to $1.48\%$, less than half that of $\m{P}^0_{\mathrm{engw}}$ ($3.79\%$) and below one fifth that of $\m{P}^0_{\m{I}}$ ($7.90\%$). 
This improvement is consistent with the construction of $\m{P}^0_{\alpha\mathrm{PL}}$: the player-specific linear corrections shift the stabilizing reference solution toward the cARE solution, placing the initial point closer to a region in which Newton-Raphson converges to an admissible equilibrium.

\subsection{Solution accuracy}
For every valid solution the residual is small and, is governed by the solver rather than by the initialization: within a given solver the converged accuracy is independent of the starting point. Policy iteration and the sequential solver reach residual sums of order $10^{-6}$--$10^{-7}$, whereas Newton--Raphson attains $10^{-14}$; the worst-case residual sum observed over all $10^4$ games never exceeds $4.5\times10^{-6}$. The proposed initialization therefore affects convergence and robustness, but not the accuracy of the returned equilibrium, which stays at the numerical floor of each solver.

\subsection{Convergence speed and computation time}
$\m{P}^0_{\alpha\mathrm{PL}}$ requires the fewest iterations for all three solvers, see~Table~\ref{tab:iter}, the reduction being most pronounced for Newton--Raphson ($4.38$ against $5.93$ and $7.44$). Since the per-iteration cost is essentially constant, the solver runtime tracks the iteration count, so $\m{P}^0_{\alpha\mathrm{PL}}$ also yields the shortest solver time. Regarding the initialization overhead itself, the identity is essentially free ($0.005$~ms), $\m{P}^0_{\mathrm{engw}}$ costs $0.666$~ms as it solves $N$ nonlinear AREs, and $\m{P}^0_{\alpha\mathrm{PL}}$ costs only $0.424$~ms, since it solves a single reference ARE followed by one linear system. 
The resulting effect (Table~\ref{tab:time}) is that $\m{P}^0_{\alpha\mathrm{PL}}$ attains the lowest total time for two of the three solvers and is competitive with the essentially cost-free $\m{P}^0_{\m{I}}$: A more informative initialization helps the solver converge fast enough to repay its modest construction cost.

\subsection{Uniqueness of the identified equilibrium}
Because the cAREs may admit several stabilizing solutions, we also examined whether the different solver--initialization combinations converge to the same equilibrium, calling two solutions identical when their value matrices agree element-wise to within $1\%$. The agreement exceeds $98\%$ in every pairing, with a worst case of $98.4\%$ indicating that the initialization influences which equilibrium is selected slightly more than the solver does. Two conclusions follow. 
First, the differences in convergence speed and robustness cannot be attributed primarily to the selection of different equilibria. Second, changing the solver or initialization appears to provide only limited scope for systematically exploring multiple equilibria.

\begin{figure*}[!t]
	\centering
\begin{tikzpicture}

\definecolor{darkgrey176}{RGB}{176,176,176}
\definecolor{darkslateblue3660152}{RGB}{36,60,152}
\definecolor{lightgrey204}{RGB}{204,204,204}
\definecolor{lightseagreen36152193}{RGB}{36,152,193}
\definecolor{mediumaquamarine142211186}{RGB}{142,211,186}

\begin{groupplot}[group style={group size=3 by 1}]
\nextgroupplot[
width=0.34\textwidth,
height=0.32\textwidth,
legend cell align={left},
legend style={
  font=\tiny,
  inner sep=1pt,
  fill opacity=0.8,
  draw opacity=1,
  text opacity=1,
  at={(0.97,0.03)},
  anchor=south east,
  draw=lightgrey204
},
tick align=outside,
tick pos=left,
title={Policy Iteration},
x grid style={darkgrey176},
xlabel={$N$},
xmin=1.45, xmax=13.55,
xtick style={color=black},
y grid style={darkgrey176},
ylabel={Mean number of iterations},
ymajorgrids,
ymin=16.6050417871222, ymax=88.1261224704336,
ytick style={color=black}
]
\addplot [semithick, mediumaquamarine142211186, mark=*, mark size=1, mark options={solid}]
table {%
2 23.6348195329087
3 35.6806722689076
4 48.5
5 52.7742303306728
6 61.0716845878136
7 65.1699029126214
8 68.4301606922126
9 73.4367245657568
10 75.7382133995037
11 80.2230769230769
12 79.1287386215865
13 81.368906455863
};
\addlegendentry{$P_{I}$}
\addplot [semithick, lightseagreen36152193, mark=*, mark size=1, mark options={solid}]
table {%
2 22.364
3 35.9927536231884
4 48.1730141458107
5 53.0421412300683
6 60.7269461077844
7 66.9709443099274
8 70.1307028360049
9 75.1377171215881
10 77.6600496277916
11 83.1052631578947
12 79.7176470588235
13 84.8751642575559
};
\addlegendentry{$P_\mathrm{engw}$}
\addplot [semithick, darkslateblue3660152, mark=*, mark size=1, mark options={solid}]
table {%
2 19.856
3 33.1854922279793
4 45.1686615886833
5 49.4510250569476
6 56.8805256869773
7 62.1973365617433
8 64.4604938271605
9 69.2704714640198
10 71.6939281288724
11 76.1820512820513
12 74.6033810143043
13 79.2369109947644
};
\addlegendentry{$P_{\alpha\mathrm{PL}}$}

\nextgroupplot[
width=0.36\textwidth,
height=0.32\textwidth,
tick align=outside,
tick pos=left,
title={Sequential Solver},
x grid style={darkgrey176},
xlabel={$N$},
xmin=1.45, xmax=13.55,
xtick style={color=black},
y grid style={darkgrey176},
ymajorgrids,
ymin=10.52445, ymax=14.89255,
ytick style={color=black}
]
\addplot [semithick, mediumaquamarine142211186, mark=*, mark size=1, mark options={solid}]
table {%
2 12.366
3 14.424
4 14.694
5 14.587
6 13.934
7 13.31
8 13.162
9 12.946
10 12.833
11 12.87
12 12.84
13 12.838
};
\addplot [semithick, lightseagreen36152193, mark=*, mark size=1, mark options={solid}]
table {%
2 11.984
3 13.839
4 13.935
5 13.4954954954955
6 12.812
7 12.058
8 11.89
9 11.612
10 11.458
11 11.444
12 11.402
13 11.359
};
\addplot [semithick, darkslateblue3660152, mark=*, mark size=1, mark options={solid}]
table {%
2 10.877
3 12.755
4 12.983
5 12.684
6 12.13
7 11.414
8 11.238
9 10.988
10 10.823
11 10.817
12 10.753
13 10.723
};

\nextgroupplot[
width=0.36\textwidth,
height=0.32\textwidth,
tick align=outside,
tick pos=left,
title={Newton-Raphson},
x grid style={darkgrey176},
xlabel={$N$},
xmin=1.45, xmax=13.55,
xtick style={color=black},
y grid style={darkgrey176},
ymajorgrids,
ymin=3.87515, ymax=8.90985,
ytick style={color=black}
]
\addplot [semithick, mediumaquamarine142211186, mark=*, mark size=1, mark options={solid}]
table {%
2 7.30260303687636
3 7.9245867768595
4 7.68756319514661
5 7.92540322580645
6 8.08625877632899
7 8.15215215215215
8 8.31731731731732
9 8.394
10 8.409
11 8.539
12 8.593
13 8.681
};
\addplot [semithick, lightseagreen36152193, mark=*, mark size=1, mark options={solid}]
table {%
2 6.01153039832285
3 6.61949685534591
4 6.46137787056367
5 6.20588235294118
6 6.23357664233577
7 5.87423312883436
8 5.92566191446029
9 5.78542510121457
10 5.66666666666667
11 5.65644171779141
12 5.71036585365854
13 5.60652395514781
};
\addplot [semithick, darkslateblue3660152, mark=*, mark size=1, mark options={solid}]
table {%
2 4.32388663967611
3 4.54141414141414
4 4.44040404040404
5 4.36418511066398
6 4.28857715430862
7 4.20520520520521
8 4.22022022022022
9 4.18
10 4.139
11 4.148
12 4.116
13 4.104
};
\end{groupplot}

\end{tikzpicture}
	\caption{Mean number of iterations until convergence versus the number of players $N$, for policy iteration (left), the sequential solver (middle) and Newton-Raphson (right). Across the whole range, the proposed initialization $\m{P}^0_{\alpha\mathrm{PL}}$ requires the fewest iterations. Note that the vertical axes differ in scale.}
	\label{fig:iter_by_N}
\end{figure*}
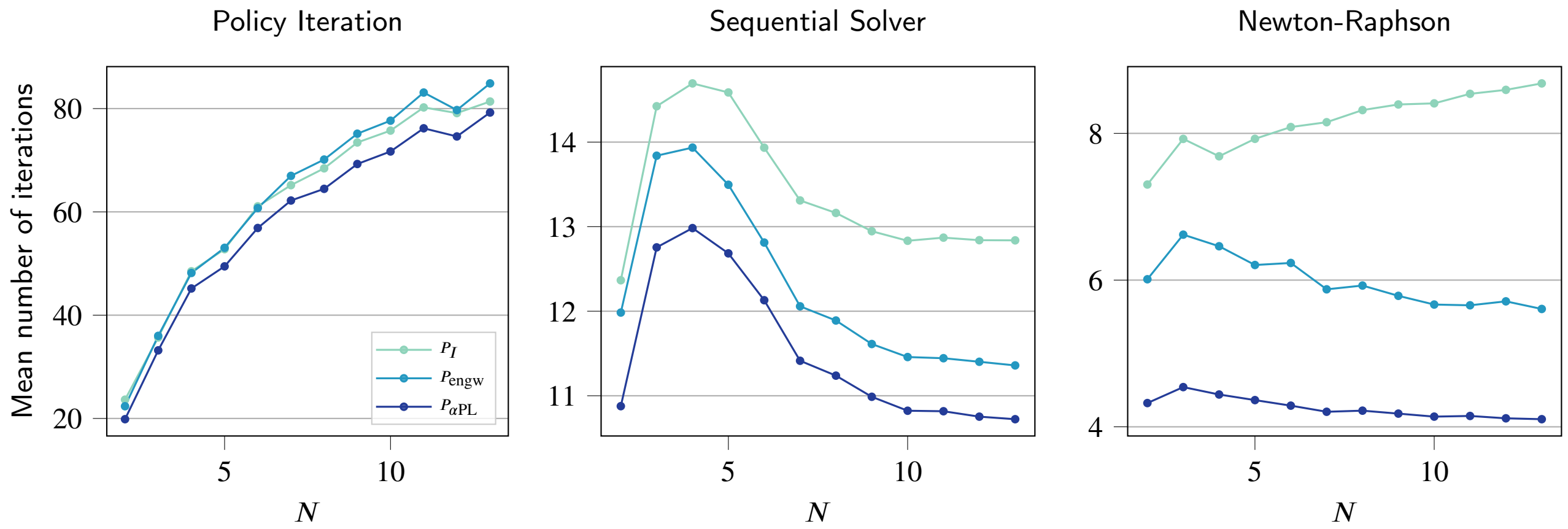

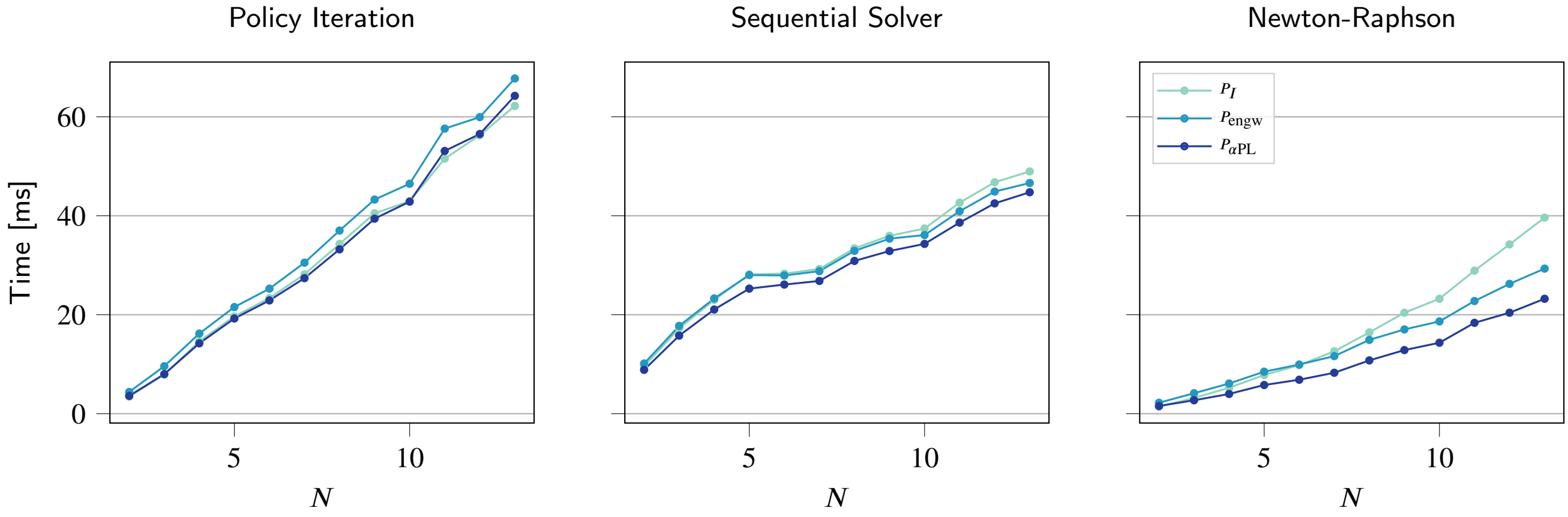
\begin{figure*}[t]
	\centering
\begin{tikzpicture}

\definecolor{darkgrey176}{RGB}{176,176,176}
\definecolor{darkslateblue3660152}{RGB}{36,60,152}
\definecolor{lightgrey204}{RGB}{204,204,204}
\definecolor{lightseagreen36152193}{RGB}{36,152,193}
\definecolor{mediumaquamarine142211186}{RGB}{142,211,186}

\begin{groupplot}[group style={group size=3 by 1}]
\nextgroupplot[
width=0.36\textwidth,
height=0.32\textwidth,
tick align=outside,
tick pos=left,
title={Policy Iteration},
x grid style={darkgrey176},
xlabel={$N$},
xmin=1.45, xmax=13.55,
xtick style={color=black},
y grid style={darkgrey176},
ylabel={Time [ms]},
ymajorgrids,
ymin=-1.89278145453492, ymax=71.0642772404198,
ytick style={color=black}
]
\addplot [semithick, mediumaquamarine142211186, mark=*, mark size=1, mark options={solid}]
table {%
2 3.46079579640667
3 7.90928492028224
4 14.6247157074323
5 19.6529349304371
6 23.3646818878229
7 28.1660951516033
8 34.3081007813273
9 40.4948644953458
10 42.9588884254185
11 51.5421018511682
12 56.2738020050379
13 62.1909648317091
};
\addplot [semithick, lightseagreen36152193, mark=*, mark size=1, mark options={solid}]
table {%
2 4.3947584881098
3 9.59497563959928
4 16.1947748051434
5 21.5523061242597
6 25.2684866193568
7 30.5200096560638
8 37.0052528201518
9 43.2769307802978
10 46.47243603699
11 57.6116445148954
12 59.9377444560268
13 67.7480472997401
};
\addplot [semithick, darkslateblue3660152, mark=*, mark size=1, mark options={solid}]
table {%
2 3.59780643507111
3 7.99598292558613
4 14.2177363024225
5 19.2436328483576
6 22.8919189343887
7 27.3918209722869
8 33.2140604335075
9 39.4110417013503
10 42.8539711388918
11 53.0981389114832
12 56.5260853721325
13 64.2358145141961
};

\nextgroupplot[
  width=0.36\textwidth,
height=0.32\textwidth,
scaled y ticks=manual:{}{\pgfmathparse{#1}},
tick align=outside,
tick pos=left,
title={Sequential Solver},
x grid style={darkgrey176},
xlabel={$N$},
xmin=1.45, xmax=13.55,
xtick style={color=black},
y grid style={darkgrey176},
ymajorgrids,
ymin=-1.89278145453492, ymax=71.0642772404198,
ytick style={color=black},
yticklabels={}
]
\addplot [semithick, mediumaquamarine142211186, mark=*, mark size=1, mark options={solid}]
table {%
2 9.58771356219222
3 17.2420688932907
4 22.9970002150003
5 28.1532353681614
6 28.3083424220749
7 29.2434053177494
8 33.4123650340844
9 35.9477185999858
10 37.4227693469438
11 42.6457988998809
12 46.7704745078299
13 48.9604298372578
};
\addplot [semithick, lightseagreen36152193, mark=*, mark size=1, mark options={solid}]
table {%
2 10.1389177192104
3 17.7401420149326
4 23.2891398290521
5 28.0129821292538
6 27.9313919389315
7 28.8035134710808
8 32.9062452561047
9 35.3752072520219
10 36.1028848668284
11 40.9147236558856
12 44.8873715110167
13 46.6285890440558
};
\addplot [semithick, darkslateblue3660152, mark=*, mark size=1, mark options={solid}]
table {%
2 8.86070917117468
3 15.7705426699249
4 21.0554912389707
5 25.2772012690039
6 26.0984677029264
7 26.8277265419601
8 30.8718613907695
9 32.8737084831446
10 34.3182916692895
11 38.6179006301027
12 42.5038340531828
13 44.7513400020616
};

\nextgroupplot[
width=0.36\textwidth,
height=0.32\textwidth,
legend cell align={left},
legend style={
  font=\tiny,
  inner sep=1pt,
  fill opacity=0.8,
  draw opacity=1,
  text opacity=1,
  at={(0.03,0.97)},
  anchor=north west,
  draw=lightgrey204
},
scaled y ticks=manual:{}{\pgfmathparse{#1}},
tick align=outside,
tick pos=left,
title={Newton-Raphson},
x grid style={darkgrey176},
xlabel={$N$},
xmin=1.45, xmax=13.55,
xtick style={color=black},
y grid style={darkgrey176},
ymajorgrids,
ymin=-1.89278145453492, ymax=71.0642772404198,
ytick style={color=black},
yticklabels={}
]
\addplot [semithick, mediumaquamarine142211186, mark=*, mark size=1, mark options={solid}]
table {%
2 1.42344848614484
3 3.17285522144006
4 5.23261672623273
5 7.82466857790056
6 9.80694467518147
7 12.6074376215523
8 16.4277952033561
9 20.3996413411369
10 23.2227790870238
11 28.9085109028529
12 34.2135117189464
13 39.6215998862026
};
\addlegendentry{$P_{I}$}
\addplot [semithick, lightseagreen36152193, mark=*, mark size=1, mark options={solid}]
table {%
2 2.21022917728804
3 4.13435651153389
4 6.10423144673541
5 8.48775255152981
6 9.93644379571315
7 11.6344862220648
8 14.9153235437471
9 17.0291274928746
10 18.6473292375053
11 22.7688784500343
12 26.2366368963008
13 29.3180811244565
};
\addlegendentry{$P_\mathrm{engw}$}
\addplot [semithick, darkslateblue3660152, mark=*, mark size=1, mark options={solid}]
table {%
2 1.55517191918996
3 2.7119970261625
4 3.98601828596609
5 5.79343597768954
6 6.86114235079463
7 8.27093103904368
8 10.7685365794452
9 12.8533189131631
10 14.3380993841347
11 18.3617346730316
12 20.4065058951819
13 23.222177814052
};
\addlegendentry{$P_{\alpha\mathrm{PL}}$}
\end{groupplot}

\end{tikzpicture}
	\caption{Mean total computation time (initialization plus solver) versus the number of players $N$, for the three solvers. For the sequential and Newton-Raphson solvers, $\m{P}^0_{\alpha\mathrm{PL}}$ (proposed) attains the lowest total time at every $N$.}
	\label{fig:time_by_N}
\end{figure*}

\subsection{Scalability}
To assess scalability, a second study generates $10^3$ instances for each $N=2,\dots,13$, with $n=m_i=2$, and compares the three solvers. Iteration trends (Fig.~\ref{fig:iter_by_N}) differ substantially with $N$: policy iteration grows monotonically, requiring roughly four times as many iterations at $N=13$ as at $N=2$; the sequential solver peaks near $N=4$ before declining toward an initialization-dependent limit; and Newton--Raphson peaks near $N=3$ for both $\m{P}^0_{\mathrm{engw}}$ and $\m{P}^0_{\alpha\mathrm{PL}}$. Across all $N$, $\m{P}^0_{\alpha\mathrm{PL}}$ requires the fewest iterations.

Solver dependence is similarly evident in invalid rates. The policy-iteration rate increases monotonically to approximately $25\%$ at $N=13$, whereas the sequential solver remains orders of magnitude lower and Newton--Raphson improves as $N$ increases. Only Newton--Raphson is noticeably sensitive to initialization: $\m{P}^0_{\alpha\mathrm{PL}}$ consistently outperforms $\m{P}^0_{\mathrm{engw}}$, keeping invalid rates below $2\%$ even at $N=2$. Accuracy is maintained throughout. While the residual sum grows approximately proportionally with $N$ for policy iteration and the sequential solver because per-player residuals accumulate, Newton--Raphson remains below $10^{-8}$ without a clear trend. All valid solutions have residuals below $10^{-6}$.

These results also identify the main limitation of the proposed initialization. As $N$ grows, constructing the reference game becomes more expensive, making the fixed per-player cost of $\m{P}^0_{\mathrm{engw}}$ marginally lower beyond approximately $N=11$. Nevertheless, the reduced iteration count from $\m{P}^0_{\alpha\mathrm{PL}}$ offsets its higher initialization cost, yielding lower total computation time across the full range considered.

\subsection{Discussion} \label{sec:discussion}
The results indicate that the proposed $\alpha$PL initialization provides a practical compromise between the cost of constructing an informed initial guess and the reduction in subsequent solver iterations. Its advantage is particularly clear for Newton-Raphson iteration, whose convergence and admissibility are most sensitive to initialization, while the sequential solver remains robust for all tested initializations. The scalability study further shows that the iteration reduction persists as the number of players increases, although the construction of the reference game becomes increasingly expensive. Since the approximation neglects second-order deviation terms, its effectiveness is expected to be strongest when the individual player costs remain sufficiently close to the aggregated reference cost. Moreover, the present formulation is restricted to decoupled control penalties; extending the construction to games with nonzero $\m{R}_{ij}$, $i\neq j$, and optimizing the aggregation weights $\theta_i$ are important directions for future work.

\section{Conclusion}\label{sec:conclusion}
We proposed an $\alpha$-potential-inspired linear approximation for initializing the cAREs of $N$-player LQ differential games with decoupled control penalties. The method requires the solution of one standard ARE and one linear system, and its cARE residual is second order in the correction magnitude. Numerically, it reduces the iteration count for all three investigated solvers and improves the robustness of the initialization-sensitive Newton-Raphson method, with the advantage persisting for $N=2,\ldots,13$. Future work will address optimized aggregation weights and conditions under which the approximation can be used directly without subsequent cARE iterations.

\printcredits

\section*{Declaration of competing interest}
The authors declare no competing interests.

\section*{Data availability}
Results are available to readers on request to the corresponding author.

\section*{Funding}
Gergely L. Kiss was supported by an Erasmus+ scholarship during his research stay at the Karlsruhe Institute of Technology (KIT). This research did not receive any other specific grant from funding agencies in the public, commercial, or not-for-profit sectors.

\section*{Declaration of generative AI and AI-assised technologies in the writing process}
During the preparation of this work the authors used ChatGPT 5.5 and Claude Opus 5.5 in order to perform language editing. After using this tool, the authors reviewed and edited the content as needed and take full responsibility for the content of the publication.


\end{document}